\documentclass[12pt]{article}
\usepackage{latexsym}
\usepackage{epsfig}

\usepackage{graphicx}
\usepackage{epstopdf}

\usepackage{epsfig,amssymb,amsmath,amscd}

\newcommand{\bq}{\begin{equation}}
\newcommand{\eq}{\end{equation}}
\newcommand{\bqa}{\begin{eqnarray}}
\newcommand{\eqa}{\end{eqnarray}}
\newcommand{\ben}{\begin{enumerate}}
\newcommand{\een}{\end{enumerate}}
\newcommand{\bc}{\begin{center}}
\newcommand{\ec}{\end{center}}
\newcommand{\bqb}{\begin{eqnarray*}}
\newcommand{\eqb}{\end{eqnarray*}}

\def\pr#1#2#3{Phys. Rev. ${\bf{#1}}$, #2 (#3)}

\def\pl#1#2#3{Phys. Lett. ${\bf{#1}}$, #2 (#3)}

\def\np#1#2#3{Nucl. Phys. ${\bf{#1}}$, #2 (#3)}

\def\jhep#1#2#3{JHEP ${\bf{#1}}$, #2 (#3)}
\def\epj#1#2#3{Eur. Phys. J. ${\bf{#1}}$, #2 (#3)}

\def\jmp#1#2#3{J. Mod. Phys. ${\bf{#1}}$, #2 (#3)}

\begin{document}
\pagenumbering{arabic}
\thispagestyle{empty}
\def\thefootnote{\fnsymbol{footnote}}
\setcounter{footnote}{1}

\begin{flushright}
Nov. 21, 2017\\
 \end{flushright}

\begin{center}
{\Large {\bf Tests of the charged Goldstone sector through inclusive
$t\bar b$ production}}.\\
 \vspace{1cm}
{\large F.M. Renard}\\
\vspace{0.2cm}
Laboratoire Univers et Particules de Montpellier,
UMR 5299\\
Universit\'{e} Montpellier II, Place Eug\`{e}ne Bataillon CC072\\
 F-34095 Montpellier Cedex 5, France.\\
\end{center}

\vspace*{1.cm}
\begin{center}
{\bf Abstract}
\end{center}

Similarly to the previously studied case of inclusive $t\bar t$ production 
we consider the inclusive $t\bar b$ production in $e^+e^-$, gluon~gluon
and photon~photon collisions. We show how it is sensitive to new physics,
connected to the charged Goldstone sector and 
possibly generated by Higgs and/or top quark compositeness.

\vspace{0.5cm}

\def\thefootnote{\arabic{footnote}}
\setcounter{footnote}{0}
\clearpage

\section{INTRODUCTION}

In a previous work, \cite{ttincl},
we had studied the specific tests of top quark 
and/or Higgs boson compositeness which could be done through the inclusive
$t\bar t +X$ production in $e^+e^-$, gluon~gluon and photon~photon collisions.\\
We had shown how the shape of $d\sigma/dM_X$ directly reflects the
presence of new states (even invisible ones) related to such compositeness.
For compositeness of the top quark and of the Higgs boson see 
\cite{partialcomp,Hcomp2,Hcomp3,Hcomp4}, for
observability of top compositeness see \cite{Tait} and for the concept of
Compositeness Standard Model (CSM), see \cite{CSMrev}.\\
This previous study of inclusive $t\bar t +X$ production
dealt with the neutral $X$ sector. New $X$ states may first consist 
of "excited" states originating
from a substructure (see for example \cite{comp})
or from the SM extension with a new sector.
In addition the basic point-like couplings may be affected by
form factors, see \cite{trcomp}, for example
originating from the spatial extension generated by the substructure.

We now extend this type of study to the charged $X$ sector. It can first involve
a new $W'$. In SM the longitudinal $W_L$ is equivalent 
to the Goldstone $G$. With the CSM concept this may also be the case
for the excited states and a  $W'_L$ equivalent to a $G'$.
It may also consist of a whole new sector generated
by a pair of subconstituents (with a total charge $\pm 1$).
We illustrate the $t\bar b X$ case. Similar effects should appear 
for $b\bar t \bar X$.
The situation is relatively simpler than in the previous neutral case
where contributions appeared from both $H$ and $Z\equiv G^0$
sectors.\\

In the following section we give illustrations 
of the inclusive $d\sigma/dM_X$ distribution
at a fixed total energy in $e^+e^-$, gluon~gluon and photon~photon collisions.
This will be done with arbitrary values of masses and couplings.
Apart from a (not shown) standard narrow  peak at $M_X=m_W$  one should 
observe a broad peak at $m_{W'}$ (or $m_{G^{'}}$) for  an "excited"
state as well as thresholds for (visible or invisible) multiparticle production.
The comparison of experimental results for this distribution with those 
of the $t\bar t +X$ case should be very instructive especially for checking the
origin of some signal possibly first seen in the neutral sector.\\

These simple illustrations may motivate further phenomenological 
and experimental studies in this respect.
For what concerns the experimental domain, 
in $e^+e^-$ collisions see \cite{Moortgat, Denterria, Craig, Englert}, 
in hadronic collisions see \cite{Contino,Richard} and in 
photon-photon collisions see \cite{gammagamma}.\\

\section{Analyses of $e^+e^-,\gamma\gamma,gg\to t\bar b  X$}

\subsection{$e^+e^- \to t\bar b  X$}

As explained in the introduction the basic SM contribution to the $M_X$ distribution
only corresponds to $X=W^-$ boson emission as shown in Fig.1. 
There are 3 ways for  $W^-$ emission: from the quark (top and bottom),
from the $Z$ and from the $e^{\pm}$ lines.

The $W^-$ can have transverse and longitudinal polarization. At high energy, in SM,
the longitudinal component $W^-_L $ is equivalent to the Goldstone $G^-$ boson. 
The corresponding diagrams are drawn in the lower level of Fig.1.\\

We now want to discuss effects of  Higgs and top quark compositeness. In the CSM spirit
the simplest class of effects corresponds to a modification of the couplings of the Higgs
and top sectors as discused in previous studies, see \cite{CSMrev}.
Illustrations have been done by
introducing effective form factors which keep the SM structure at low energies,
like
\bq
F(s)={s_0+M^2\over s+M^2}~~\label{FF}
\eq
where $s_0$ is a threshold and $M$ a new physics scale taken for example equal
to $1$ TeV.
See also \cite{trcomp} where the corresponding concept of effective mass has 
been introduced. 
In this paper we will use this expression for showing the change produced in the inclusive
distribution when such a form factor affects the Goldstone couplings,
replacing $s$ by $M^2_X$ in the above equation.\\

But several types of additional new terms may also appear.
A first type may consist of the occurence of "excited" states.
One may expect the occurence of an "excited" $W'$, with, in the CSM spirit, the $W'_L$
component equivalent to an "excited" Goldstone state $G'^-$.
We illustrate its presence with a mass $M_{G'}=M_{W'}=1$ TeV, 
a width $\Gamma_{G'}=\Gamma_{W'}=0.1$ GeV and couplings to $W$ and top quark similar 
to those of the standard $G$.
The diagrams for $G'$ production are taken similar to those for the 
standard $G$ in Fig.1.\\

In the next step we assume the existence of a new  (visible or partly invisible) sector
coupled to the Goldstone or directly to the top quark.
It may be a strongly coupled sector or the result of a substructure which
creates multibody production similarly to the case of hadronic production generated
after quark+antiquark creation.
As an example we will assume that, after their production
according to the diagrams of Fig.1, the Goldstone boson
creates a pair of subconstituents with a mass $m_0=1$ TeV. Multiparticle production 
then occurs automatically with a threshold at $M_X=2m_0=2$ TeV.\\

We have computed the total $M_X$ distribution of the inclusive $t\bar b$ production,
$d\sigma/dM_X$, by integrating the differential cross section over the energies 
and angles of the $t$ and the $\bar b$ with $M_X^2=(q-p_t-p_{\bar b})^2$ where $q$
is the total center of mass momentum, $s=q^2$.\\

Arbitray cuts have been used in order to avoid collinear singlularities.  The
absolute values of the cross sections have no predictive meaning, we only want to
discuss shapes which may be typical of the various above dynamical assumptions.

These shapes of the $M_X$ distribution are shown in Fig.4 for a total center 
of mass energy of $\sqrt{s}=5$ TeV.
The basic SM shape is only due to $W^-$ boson emission
(the trivial peak at $m_{W}$ is not shown).
New effects appear clearly with a
bump at $m_{G'}$ and a multiparticle threshold at $2m_0$.\\

We also show how these shapes would
be modified by the presence of an effective form factor $F(M^2_X)$ affecting all the 
Goldstone couplings (basic and excited).\\

To appreciate these new effects in an other way we illustrate at the lower level of Fig.4
the corresponding ratios of the total new contributions (SM+$G$ sector)
over the pure SM one.\\

\subsection{$gluon~gluon \to t\bar b  X$}

The lowest order SM diagrams are shown in Fig.2
(gluon gluon symmetrization should be understood). The $W^-$ 
can only be emitted by the final top and bottom lines (see upper level of Fig.2).  
Contrarily to the $e^+e^-$  case there is no emission from the initial lines at this order.\\
At this step of our study we ignore the QCD corrections and in particular the gluon emission.\\
In SM longitudinal $W^-_L$ is equivalent to $G^-$ whose production
is described in the lower level of Fig.2.\\
We then add the contributions of an excited 
$G^{'-}$ state and the continuum of multiparticle
production from the Goldstone sector.\\
We also consider the effects of a form factor affecting the
$G^-$ couplings as well as those of the excited state.\\ 
The corresponding illustrations are given in Fig.5 for $\sqrt{s}=15$ TeV.
Bump and threshold effects are qualitatively similar to those of 
the $e^+e^-$ case except for
the constant ratio for high $M_X$ due to the dominance of the continuum
production parametrized with the same coupling as the one of the SM Goldstone.
This is obviously arbitrary and could be different with another 
portal dynamics.\\

\subsection{$\gamma\gamma \to t\bar b  X$}

The corresponding SM diagrams are shown in Fig.3
(with photon photon symmetrization).  The new feature is that the $W^-$ 
can also be emitted by the initial photon lines.\\
$G^-$ equivalent to longitudinal $W^-_L$ is produced according to the lower level 
set of diagrams of Fig.3.\\
New contributions from an excited state and multibody production created in the
Goldstone sector generate bump and threshold effects shown in Fig.6. The results 
are rather similar to those of the $gluon~gluon \to t\bar b  X$ process.
Although the size of the cross sections are larger (because of the
additional diagrams) the ratios are very similar
even for the form factor effects (lower level of Fig.6).\\

\section{Conclusion}

We have shown that the inclusive $t\bar b$ production in 
$e^+e^-$, gluon~gluon and photon~photon  collisions may give
remarkable signals of new physics properties 
connected through the charged Goldstone sector.\\
We have illustrated the possible effects of an "excited" $W'$ (or $G^{'}$) state
and of (parton like) multiparticle production above a threshold
originating from some substructure. The possible modifications of the point-like
Goldstone couplings by form factors are also considered.\\
Such an inclusive study should be especially fruitful if this
multiparticle production consists, at least partly, of unvisible particles.\\
This search should be complementary to the one concerning the
neutral sector done through the inclusive $t\bar t$ production.\\ 
In case departures with respect to SM expectations were
observed it would be interesting to check
if there are relations between the signals in the charged and in the
neutral sectors. They may give indications about the origin of these
effects and the underlyig dynamics. 
As already mentioned in \cite{ttincl} confirmations should also be searched
with detailed analyses of exclusive Higgs and Goldstone production processes.\\

In this first exploration we ignored electroweak and QCD radiative corrections;
we just wanted to explore what gross new features would be generated
by such compositeness assumptions. Further phenomenological 
and experimental studies may be done in this respect.\\

\newpage

\begin{figure}[p]
\[
\epsfig{file=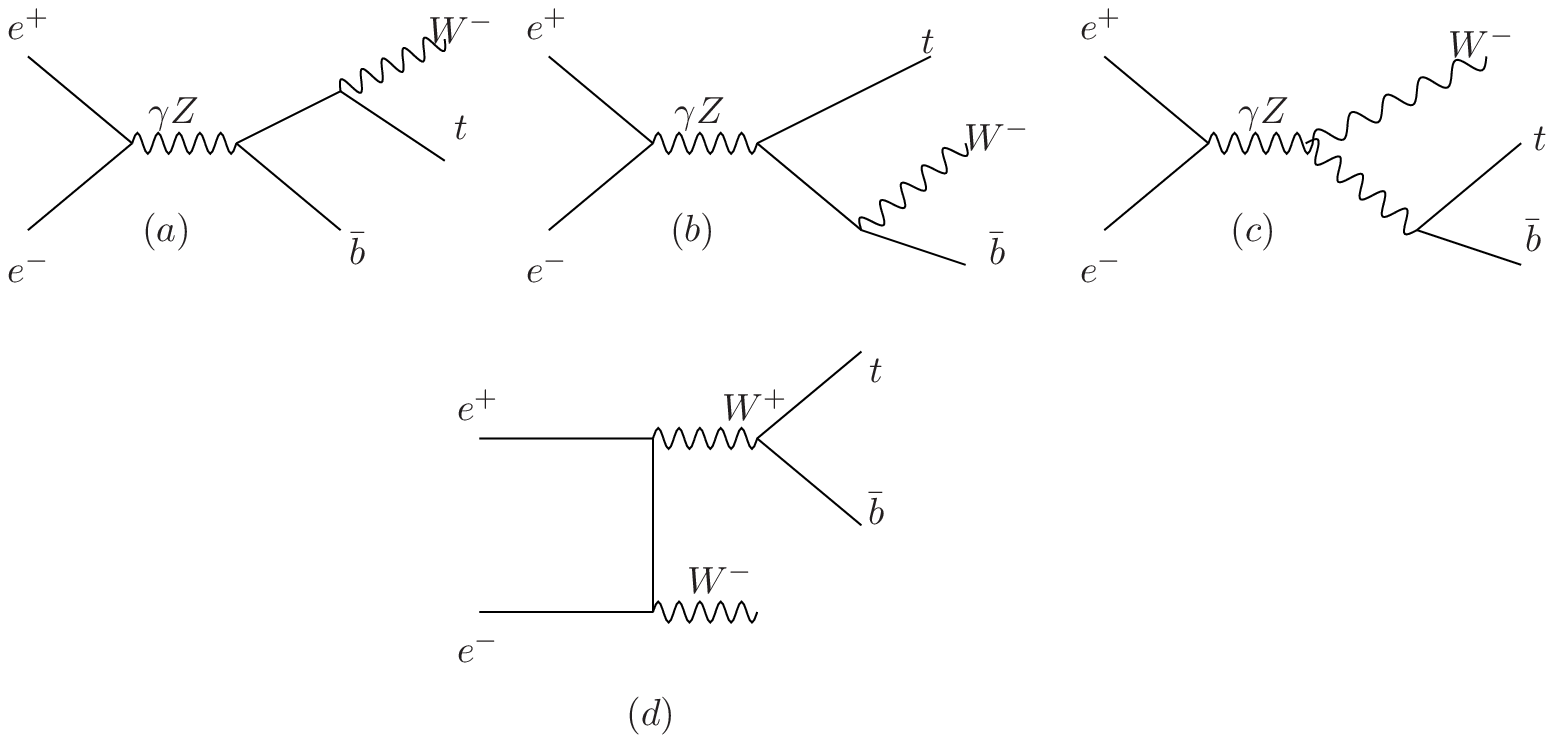 , height=8.cm}
\]\\
\[
\epsfig{file=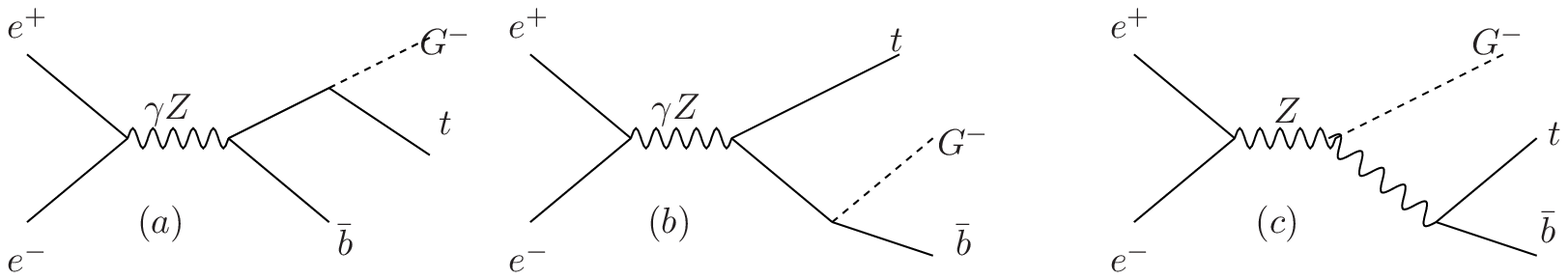 , height=3.cm}
\]\\
\caption[1] {Lowest order SM diagrams for $e^+e^- \to t\bar b  W$
and $e^+e^- \to t\bar b  G^-$ . In both (c) diagrams the internal waving line
corresponds to $W^+$ and $G^+$ contributions.}
\end{figure}

\clearpage

\begin{figure}[p]
\[
\epsfig{file=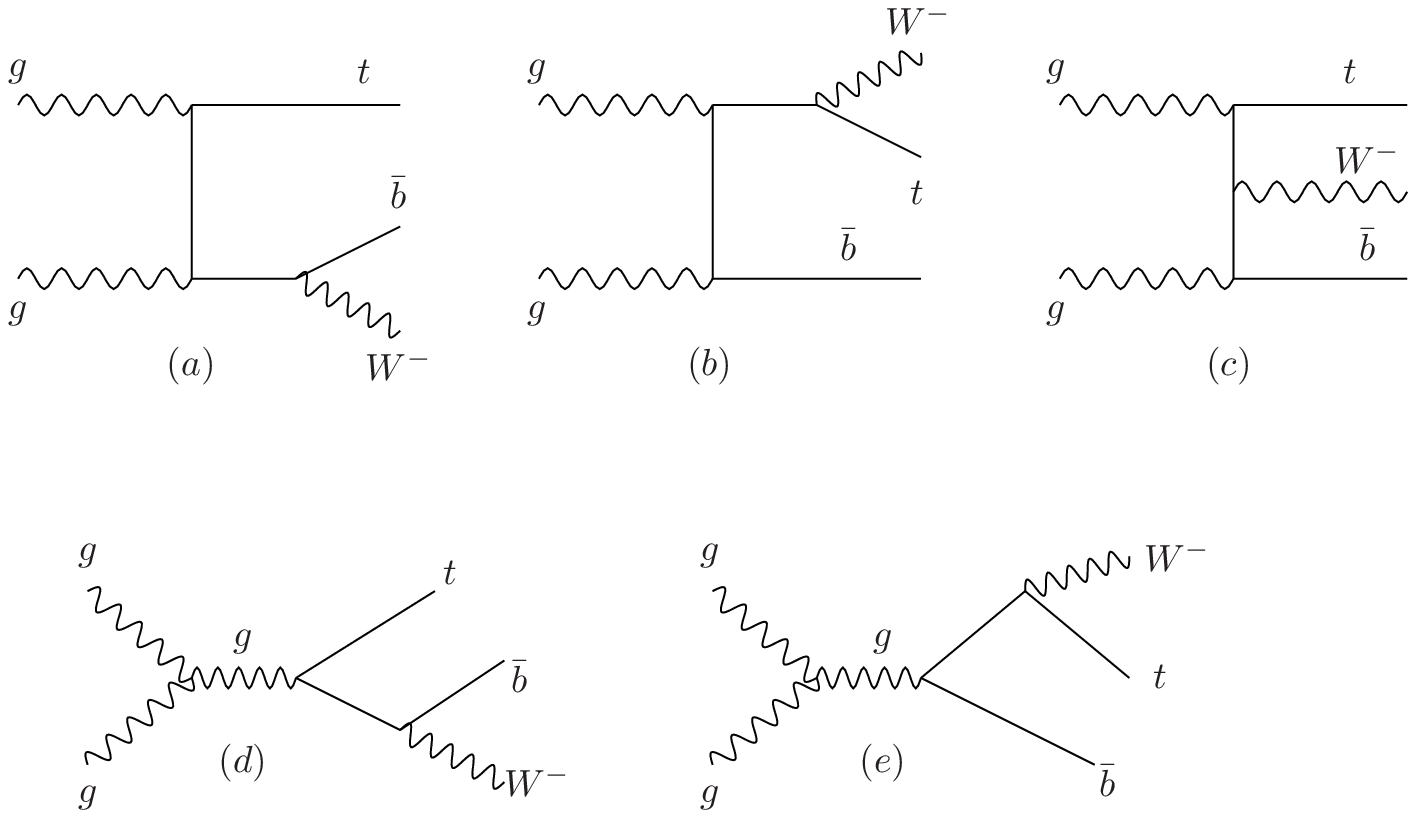 , height=10.cm}
\]\\
\[
\epsfig{file=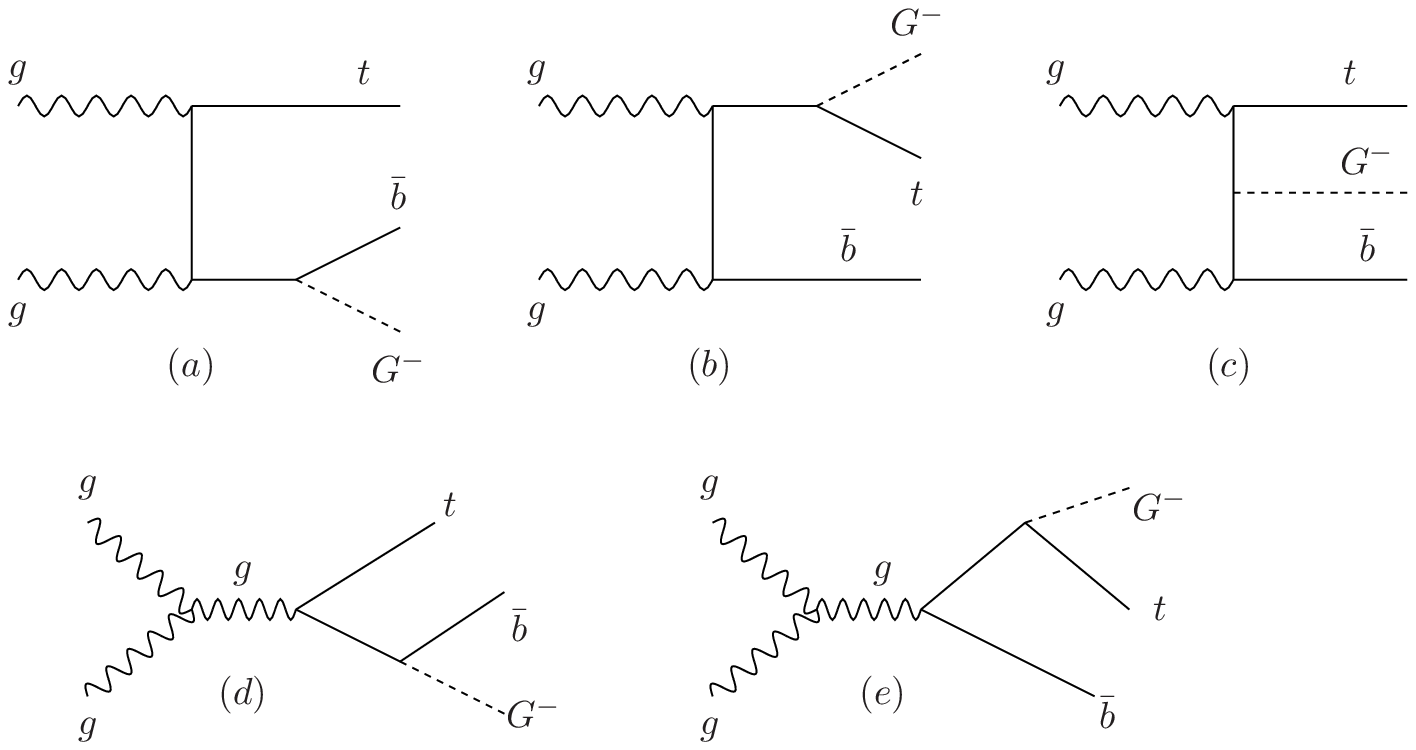 , height=8.cm}
\]\\
\caption[1] {Lowest order SM diagrams for $gluon~gluon\to t\bar b  W^-$
and $gluon~gluon\to t\bar b  G^-$.}
\end{figure}

\clearpage

\begin{figure}[p]
\[
\epsfig{file=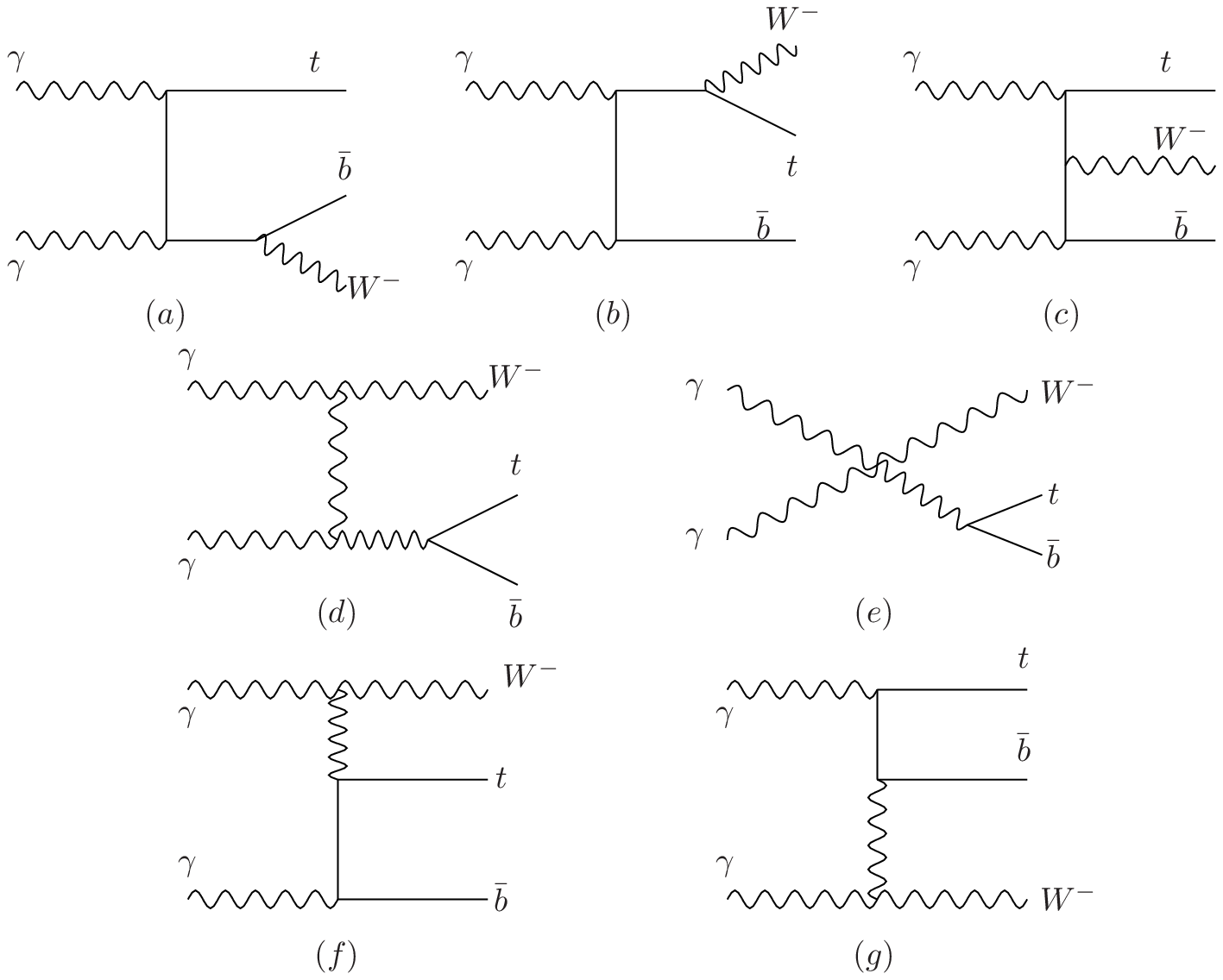 , height=12.cm}
\]\\
\vspace{-1cm}
\[
\epsfig{file=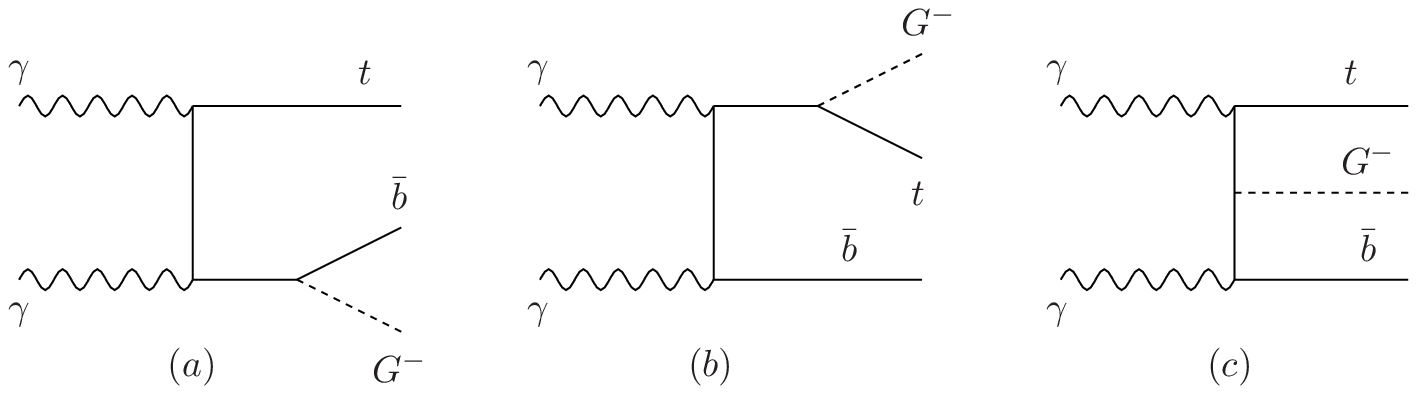 , height=4.cm}
\]\\
\vspace{-3cm}
\[
\epsfig{file=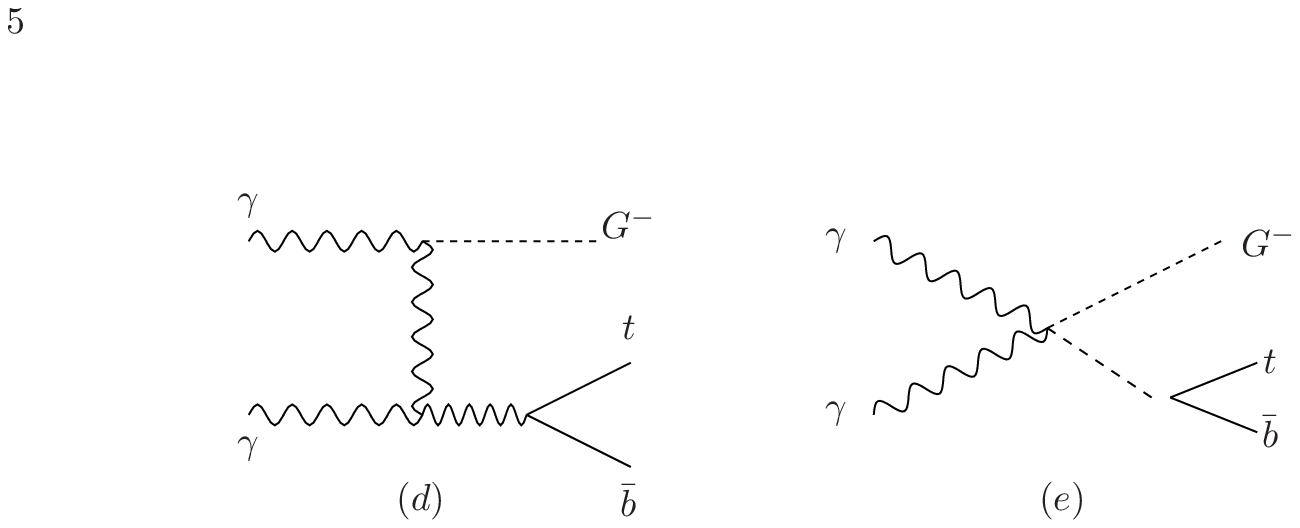 , height=5.cm}
\]\\
\vspace{-1cm}
\caption[1] {Lowest order SM diagrams for $\gamma\gamma\to t\bar b  W^-$
and $\gamma\gamma\to t\bar b  G^-$. In both (d) diagrams the internal waving lines
corresponds to $W^+$ and $G^+$ contributions.}
\end{figure}

\clearpage

\begin{figure}[p]
\[
\epsfig{file=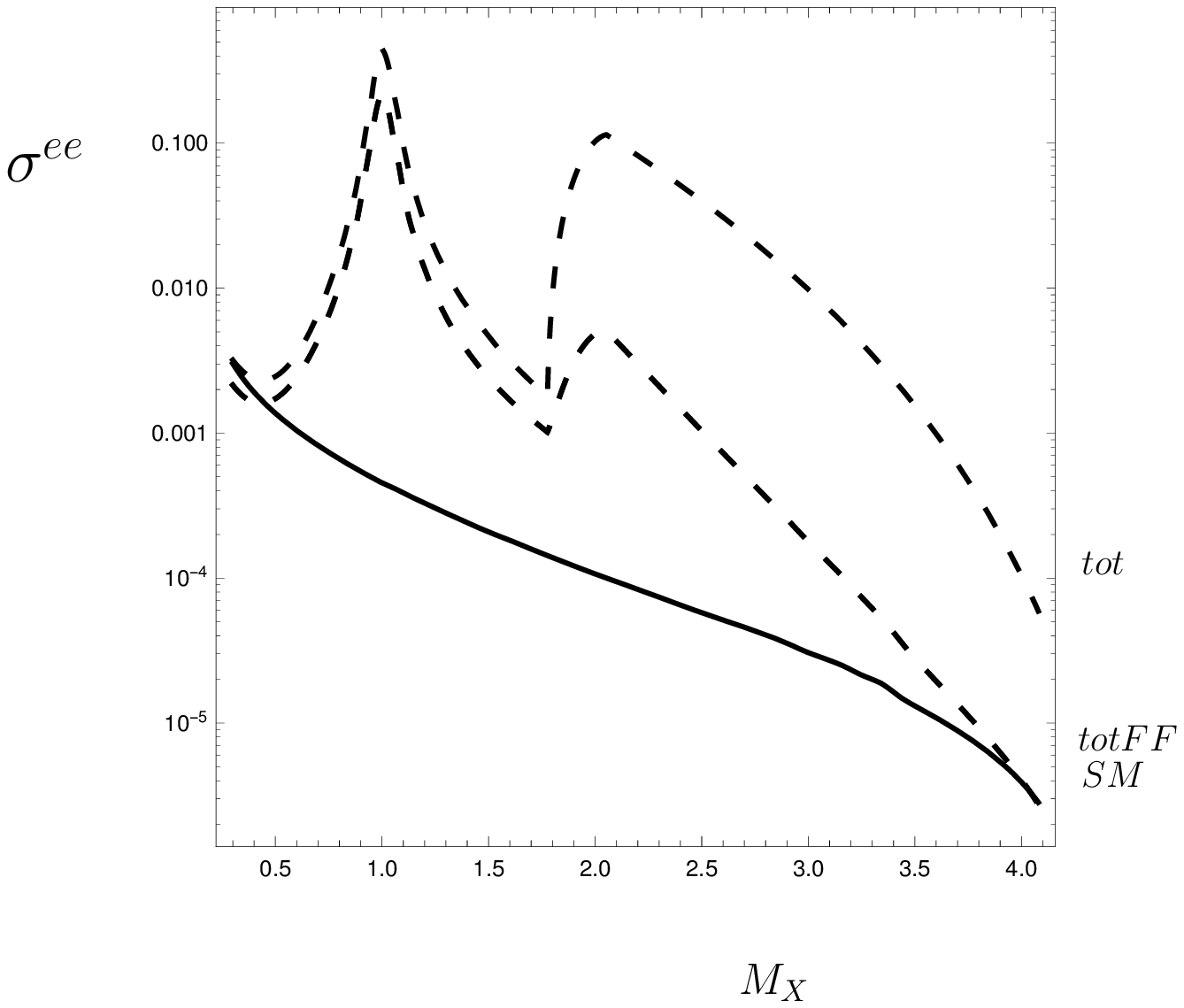, height=7.cm}
\]\\
\[
\epsfig{file=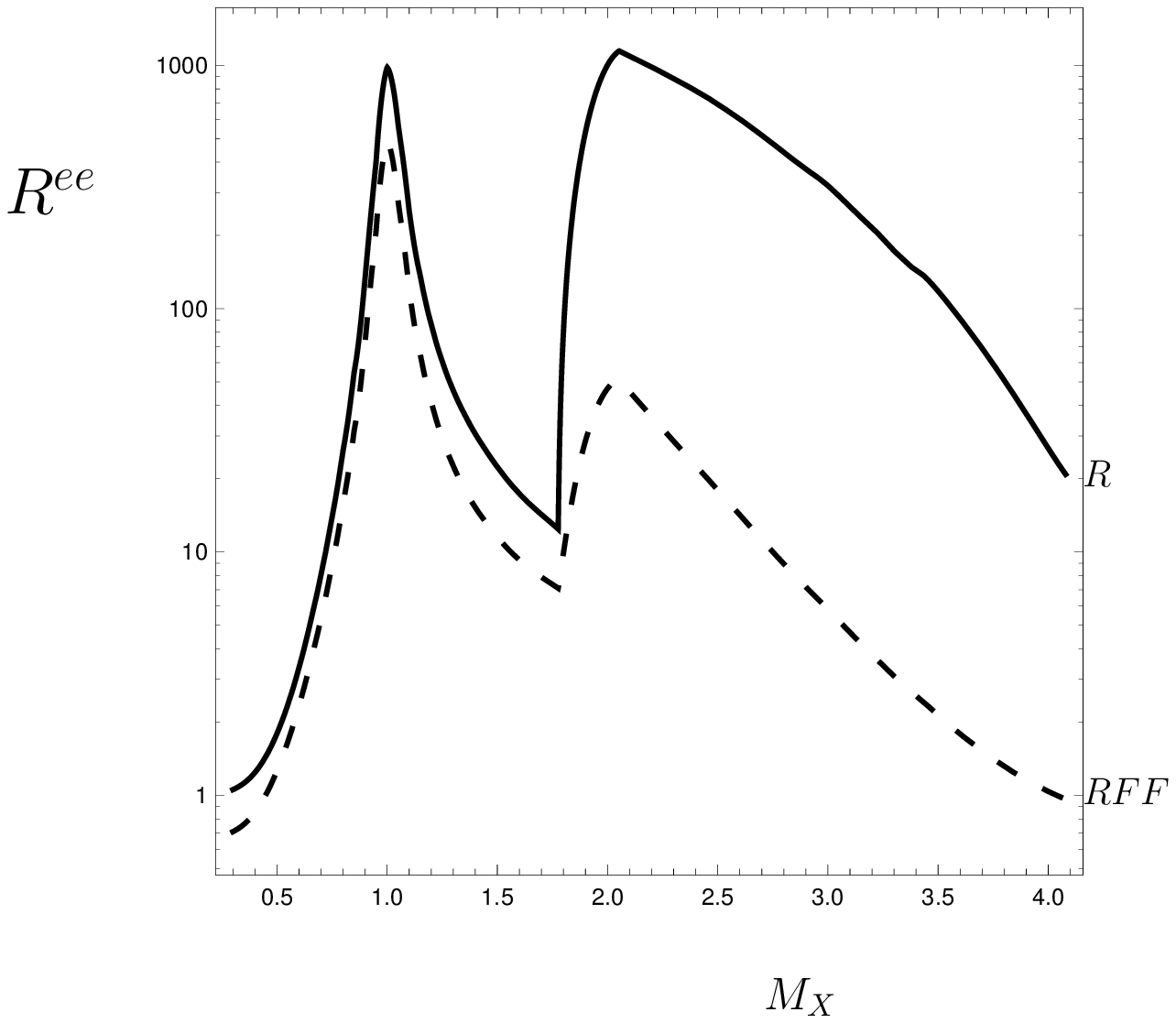, height=7.cm}
\]\\
\caption[1] {$M_X$ distribution of the $e^+e^-\to t\bar b  X$ cross section
(upper level) and ratio of the new value over the SM one (lower level); the
total new case is drawn without and with form factor effect.}
\end{figure}

\begin{figure}[p]
\[
\epsfig{file=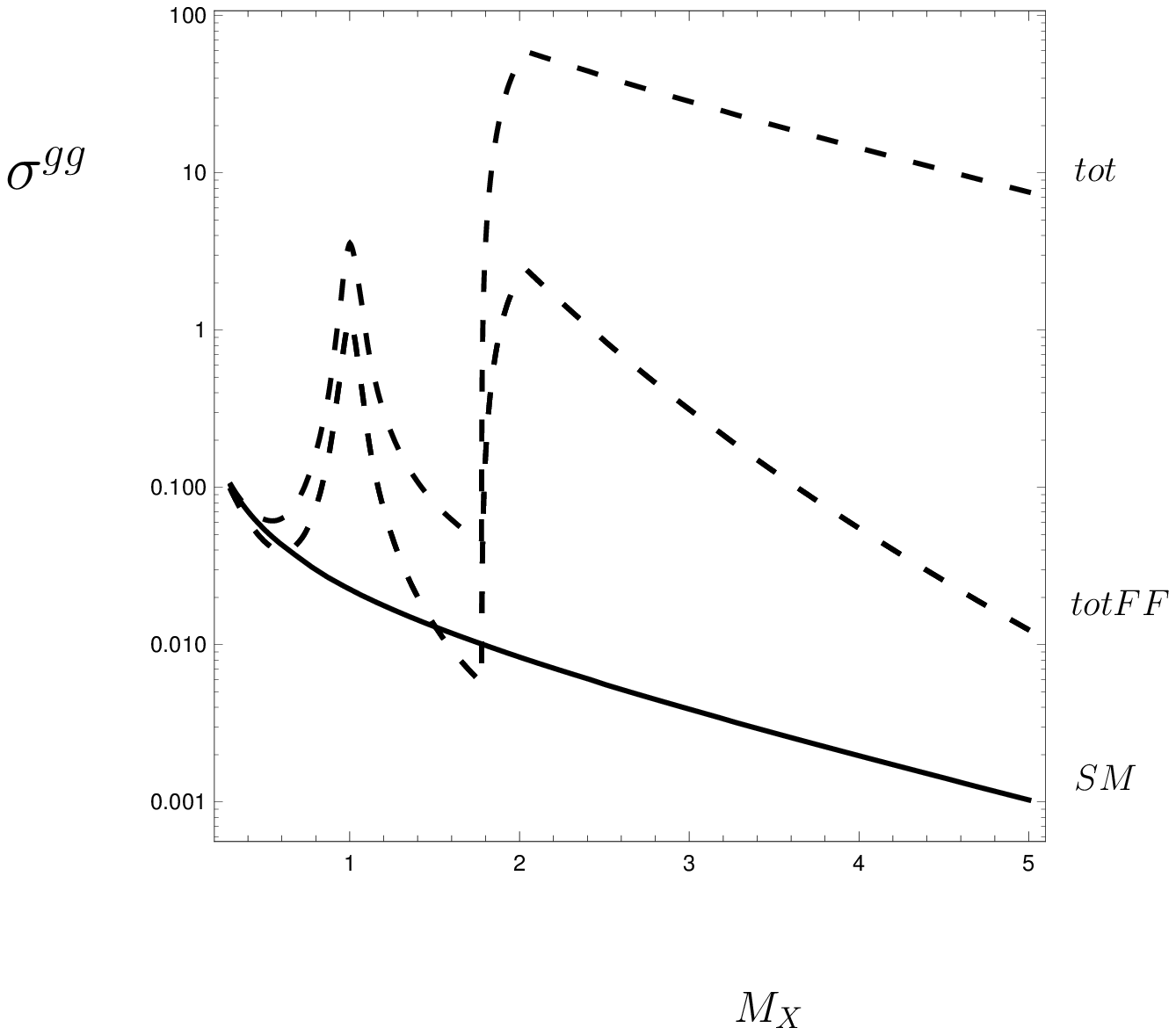, height=8.cm}
\]\\
\[
\epsfig{file=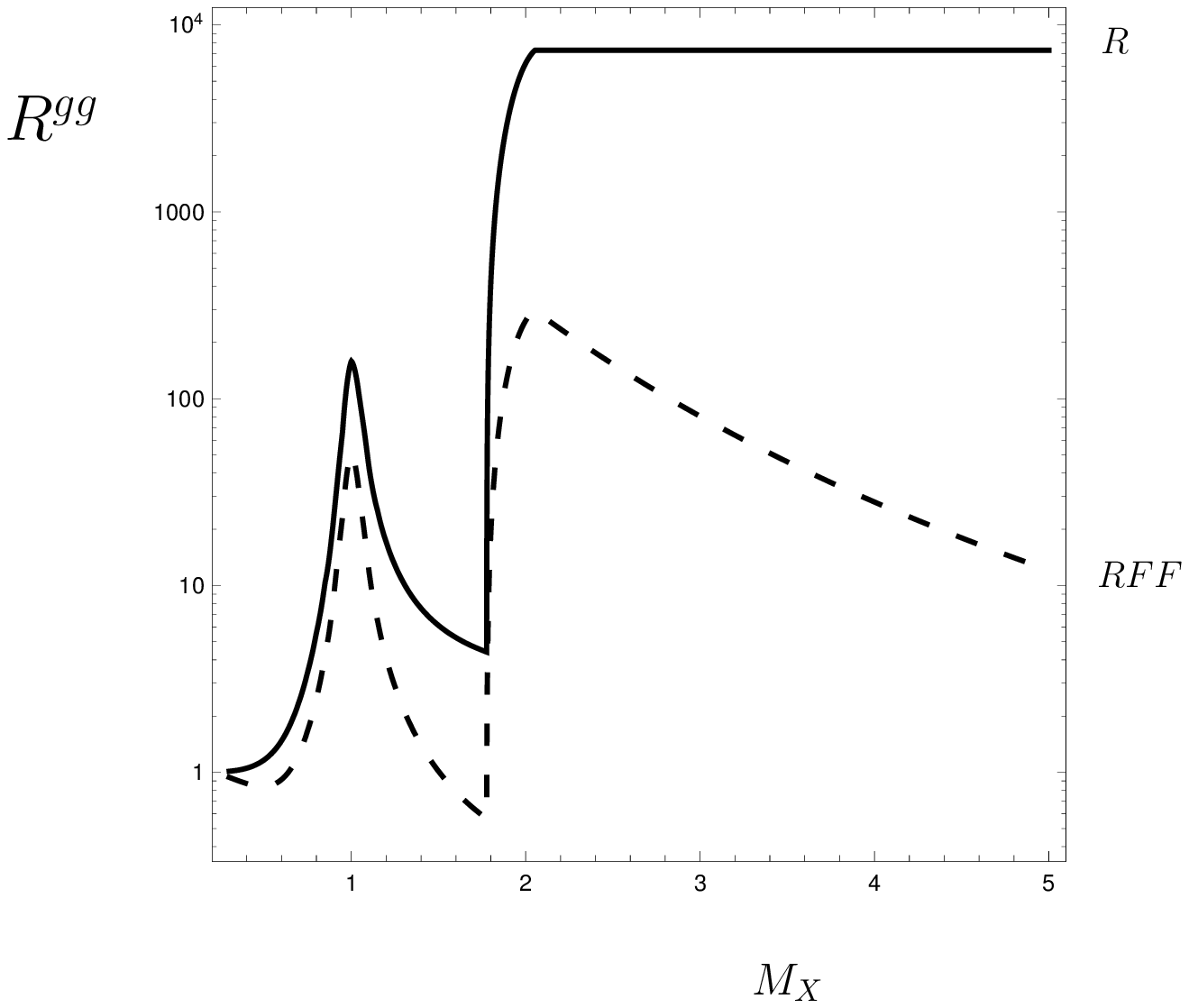, height=8.cm}
\]\\
\caption[1]  {$M_X$ distribution of the $gluon~gluon\to t\bar b  X$ cross section
(upper level) and ratio of the new value over the SM one (lower level); the
total new case is drawn without and with form factor effect.}
\end{figure}
\clearpage

\begin{figure}[p]
\[
\epsfig{file=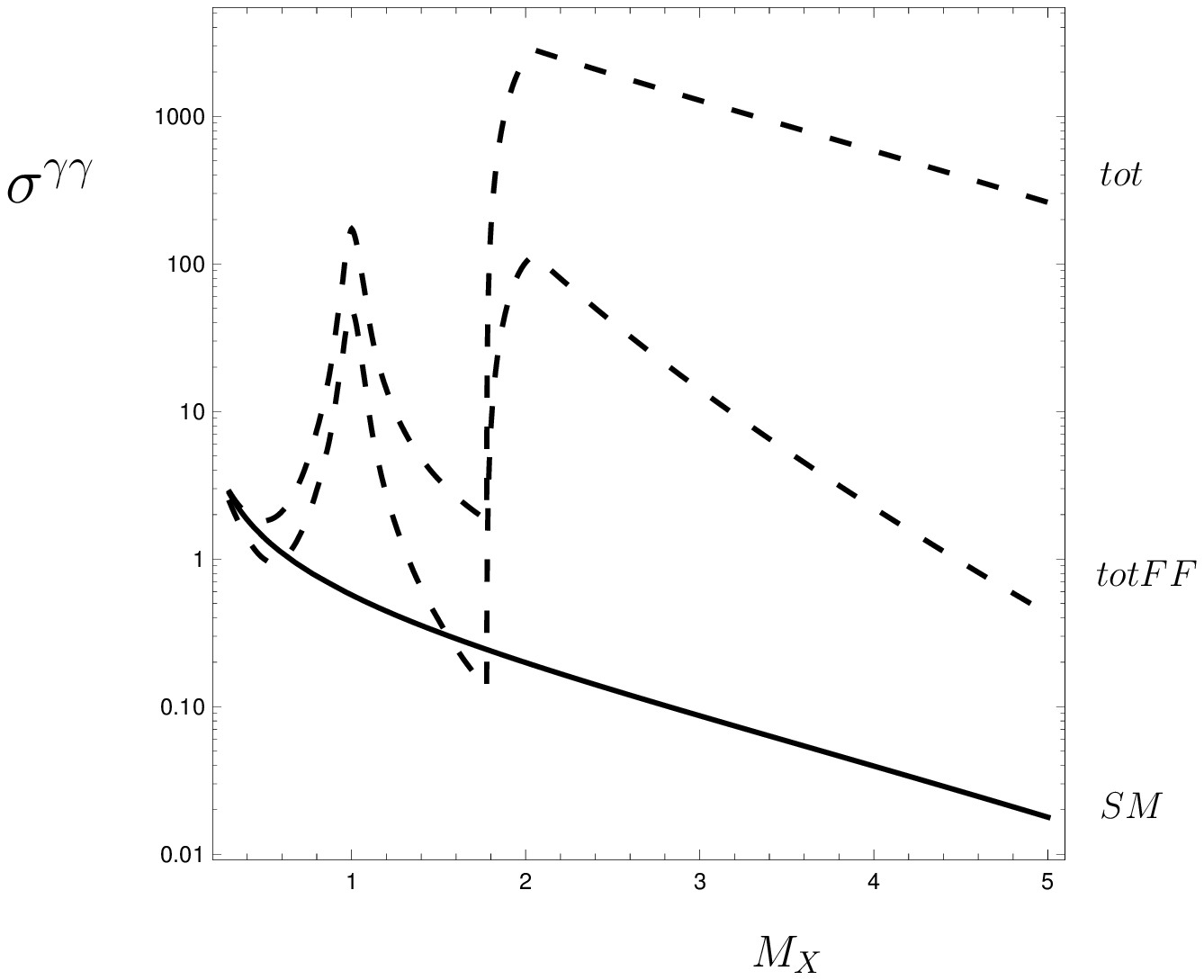, height=8.cm}
\]\\
\vspace{-1cm}
\[
\epsfig{file=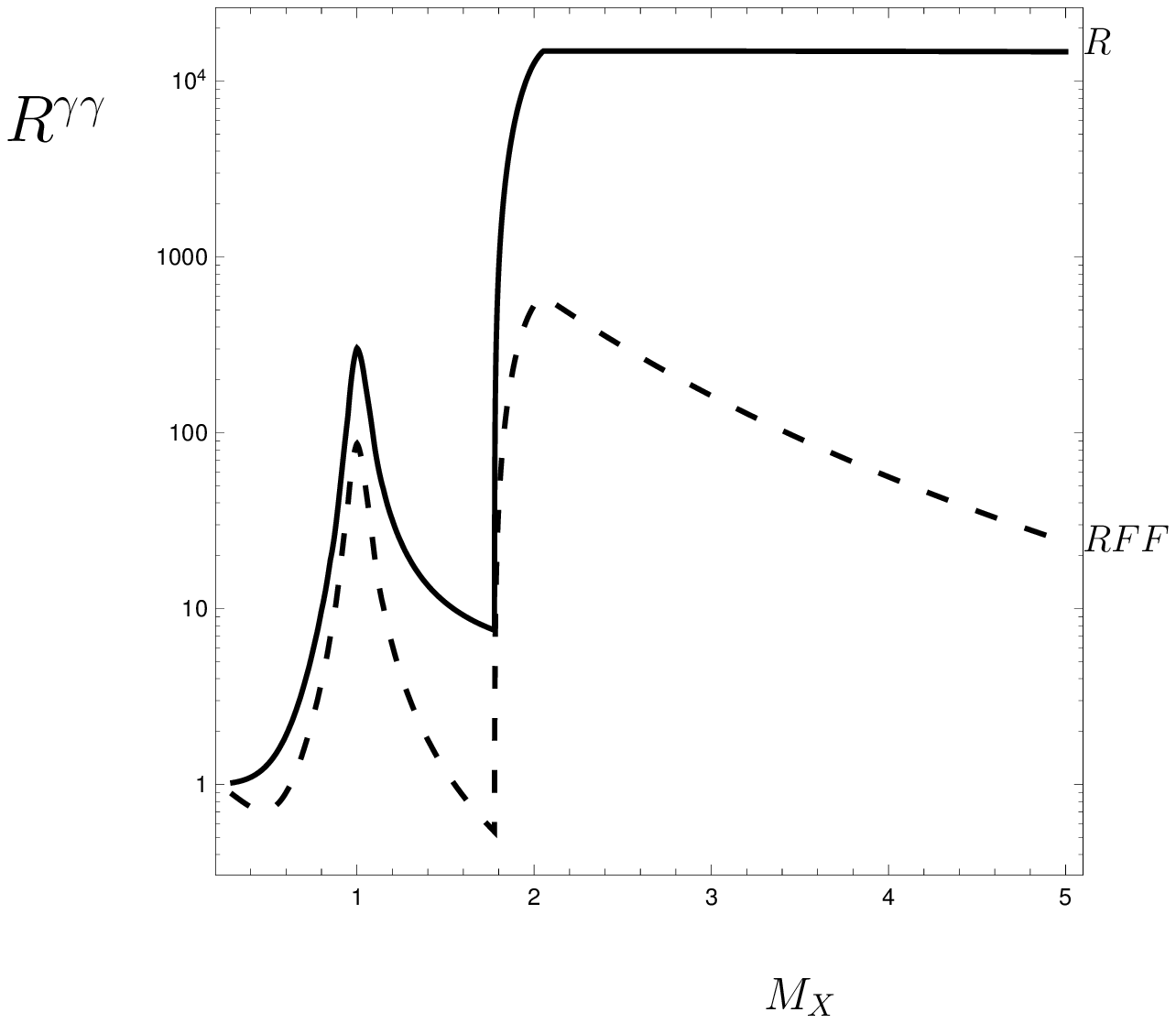, height=8.cm}
\]\\
\caption[1]  {$M_X$ distribution of the $\gamma\gamma\to t\bar b  X$ cross section
(upper level) and ratio of the new value over the SM one (lower level); the
total new case is drawn without and with form factor effect.}
\end{figure}

\end{document}